\documentclass[aps,prb,twocolumn,superscriptaddress,showpacs,amsmath,amsfonts,amssymb,floatfix]{revtex4}
\usepackage{amsmath,amsfonts,amssymb,dsfont,color,bbm,tikz,bbold}
\usepackage{graphicx}
\usepackage{dcolumn}
\usepackage{bm}
\usepackage{epstopdf}
\usepackage{epsfig}
\usepackage{environ}
\NewEnviron{myequation}{%
\begin{equation}
\scalebox{0.85}{$\BODY$}
\end{equation}
}

\begin{document}
\title{Floquet Semimetal with Floquet-band Holonomy}

\author{Longwen Zhou}
\affiliation{Department of Physics, National University of Singapore, Singapore 117546}
\author{Chong Chen}
\affiliation{Center for Interdisciplinary Studies and Key Laboratory for Magnetism and
Magnetic Materials of the MoE, Lanzhou University, Lanzhou 730000, China}
\author{Jiangbin Gong} \email{phygj@nus.edu.sg}
\affiliation{Department of Physics, National University of Singapore, Singapore 117546}


\begin{abstract}
Exotic topological states of matter such as Floquet topological insulator or Floquet Weyl semimetal can be induced by periodic driving. This work proposes
a Floquet semimetal with Floquet-band holonomy. That is, the system is gapless, but
as a periodic parameter viewed as quasimomentum of a synthetic dimension completes an adiabatic cycle, each Floquet band as a whole exchanges with other Floquet bands. The dynamical manifestations of such Floquet-band holonomy are studied. Under open boundary conditions,
we discover anomalous chiral edge modes localized only at one edge, winding around the entire quasienergy Brillouin zone, well separated from bulk states, and  possessing
holonomy different from bulk states.
These remarkable properties
are further exploited to realize quantized or half-quantized edge state pumping.
\end{abstract}
\pacs{03.65.Vf, 05.60.Gg, 05.30.Rt, 73.20.At}


\maketitle

\section{Introduction}
It is more recognized than ever that
exotic topological states of matter can be induced by periodic driving~\cite{oldFTPs}.
The resulting nonequilibrium phases, now termed as Floquet
topological phases~(FTPs), include Floquet topological insulator~\cite{OkaPRB2009,InouePRL2010,LindnerNatPhys2011,GuPRL2011,DerekPRL2012,
DelplacePRB2013,KatanPRL2013,UsajPRB2014,TorresPRL2014,DehghaniPRB2015,TitumPRL2015,SedrakyanPRL2015,FarrellPRL2015,CarpentierPRL2015,BandresPRX2016}, Floquet Weyl semimetal~\cite{BomantaraPRE2016,WangEPL2014,WangPRB2016}, etc.
The features of FTPs may go well beyond those familiar topological phases
found in static systems \cite{DerekPRB2014,TongPRB2013,SkirloPRL2015,Xiong2016}.
FTPs can also host fascinating edge modes~\cite{JiangPRL2011,LababidiPRL2014,RudnerPRX2013} not seen before, thus challenging our understanding of
bulk-edge correspondence in topological matter. Proof-of-principle
experimental studies of FTPs have also attracted great interest~\cite{RechtsmanNat2013,WangScience2013,HafeziNatPhots2013,Chong2015,Szameit16}.

Let $U$ be the one-period~($T$) propagator of a periodically driven system. If $U$ has an eigenvalue
$\exp(-{\rm i}\phi)$, the associated quasienergy is $E=\frac{\phi}{T}$~($\hbar=1$ throughout). One key reason why FTPs can accommodate previously unknown physics
is that $E$ is defined only up to a Brillouin zone $[-\frac{\pi}{T}, \frac{\pi}{T})$. That is, $E$ itself is a periodic variable and can wind around the Brillouin zone.
Indeed, precisely because of this,
chiral edge states may coexist with
topologically trivial bulk states~\cite{RudnerPRX2013}, and anomalous chiral edge states
with opposite group velocities may emerge at the same edge in Floquet Chern
insulators~\cite{DerekPRB2014,LababidiPRL2014}. The recent discovery
of ``anomalous Floquet Anderson insulator'' (AFAI)~\cite{TitumPRX2016} is again
understood in terms of quasienergy winding~(plus localization of bulk states).

In addition to quasienergy winding, periodically driven systems may possess
exotic holonomy~\cite{TanakaPRL2007,MiyamotoPRA2007}. Let $U(\lambda)$ describe a class of Floquet propagators continuously parameterized by $\lambda$, with $U(\lambda)=U(\lambda+2\pi)$.
With Floquet holonomy, the system emanating from one eigenstate
of $U(\lambda)$ may adiabatically evolve to another orthogonal state as $\lambda$ slowly increases by $2\pi$. That is, eigenstates of $U(\lambda)$ may have made transitions even though
$U(\lambda)$ itself has returned.  This remarkable phenomenon~(much different from
Berry phase or Wilczek-Zee phase) is still under investigation ~\cite{CheonEPL2009,TanakaAP2009,TanakaPRE2014}.
How such Floquet holonomy may lead to novel FTPs is the main motivation of this paper.

We start with an intriguing observation that each individual Floquet band as a whole
can exchange with other Floquet bands if we tune $\lambda\rightarrow \lambda+2\pi$ adiabatically. Such a behavior, called Floquet-band holonomy below, brings about a novel
Floquet semimetal with vanishing indirect band gaps. Due to the Floquet-band holonomy,
bulk-state geometrical charge pumping varies from cycle to cycle
and induces population transfer within one unit cell.
Using a simple model system under open boundary conditions, we also discover anomalous chiral edge modes localized only at one edge. They may wind around the entire quasienergy Brilloune zone and show holonomy behavior different from bulk states.  This indicates a new face of bulk-edge correspondence.  We further exploit the clear separation of the found anomalous edge modes from the bulk states to realize quantized or half-quantized edge state pumping.

\section{A periodically quenched lattice model}
Consider noninteracting particles hopping on a one-dimensional lattice. Each unit cell contains three sublattice sites
A, B and C. Subject to periodic quenching~($T=1$ in dimensionless units), the system is described
by a piecewise constant Hamiltonian:
\begin{equation}
\hat{H}(t)=\begin{cases}
\hat{H}_{1} & 0\leq t<\frac{1}{2}\\
\begin{aligned}\hat{H}_{2}\end{aligned}
 & \frac{1}{2}\leq t<1,
\end{cases}
\label{eq:Lattice_H}
\end{equation}
%
where
\begin{equation}
\hat{H}_{1}\equiv2\lambda\sum_{n}\hat{a}_{n}^{\dagger}\hat{a}_{n}
\end{equation}
 and
\begin{eqnarray}
\hat{H}_{2}&=&\sqrt{2}(J+V)\sum_{n}(\hat{a}_{n}^{\dagger}\hat{b}_{n}+\hat{b}_{n}^{\dagger}\hat{c}_{n}+{\rm h.c.}) \nonumber \\
&& + \sqrt{2}(J-V)\sum_{n}(\hat{b}_{n}^{\dagger}\hat{a}_{n+1}+\hat{c}_{n}^{\dagger}\hat{b}_{n+1}+{\rm h.c.}).
\end{eqnarray}
$\hat{a}_{n}^{\dagger}/\hat{a}_{n}$, $\hat{b}_{n}^{\dagger}/\hat{b}_{n}$,
$\hat{c}_{n}^{\dagger}/\hat{c}_{n}$ are creation/annihilation operators
in A, B, and C sublattices in unit cell $n$. For the first half period, only an on-site potential of strength $2\lambda$ acts on
sublattice A. Then the Hamiltonian is quenched to $\hat{H}_2$, which extends
the Su-Schrieffer-Heeger~(SSH) model~\cite{SuPRL1979} to a trimerization
situation. $\sqrt{2}(J+V)$ and $\sqrt{2}(J-V)$
can hence be understood as intracell and intercell hopping amplitudes, with their
relative strength modulated by $V$. $\hat{H}_{2}$ may be simulated by cold atoms
in optical lattices~\cite{RuostekoskiPRL2002,JavanainenPRL2003,RuostekoskiPRA2008,LiuXPRL2013,LiNatCom2013,ZhangPRA2015}.

\begin{figure} 
\begin{center}
\resizebox{0.35\textwidth}{!}{%
  \includegraphics[trim=0.6cm 0.6cm 0.3cm 10.4cm, clip=true, height=!,width=15cm]{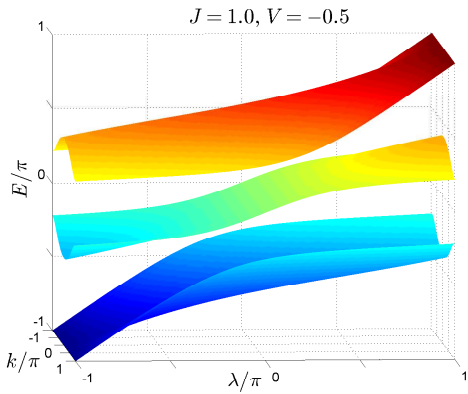}
}
\end{center}
\caption{(color online) Quasienergy spectrum of a Floquet semimetal, with $J=1$ and $V=-0.5$.
Three Floquet bands can undergo interchange as $\lambda\rightarrow \lambda+2\pi$ adiabatically.}
\label{Fig-Bulk-Spectrum}
\end{figure}

The Floquet operator is simply
\begin{equation}
\hat{U}(\lambda) = e^{-i\frac{\hat{H}_{2}}{2}}
 e^{-i\lambda \sum_{n}\hat{a}_{n}^{\dagger}\hat{a}_{n}}.
\end{equation}
Though $\hat{H}(t)$ is not periodic in $\lambda$, $\hat{U}(\lambda)$ is apparently
periodic in $\lambda$ with period $2\pi$, regardless of the boundary condition
for the lattice dimension.
One can then interpret $\lambda$ as the quasimomentum along
a synthetic dimension perpendicular to the lattice dimension.
Experimentally, a synthetic dimension may be simulated by coupling
internal states of cold atoms with Raman beams~\cite{BoadaPRL2012,CeliPRL2014,ZengPRL2015,ManciniSci2015,StuhlSci2015,PricePRL2015,OzawaPRA2016}.


\section{Floquet-band holonomy}
Consider now the periodically quenched
lattice under periodic boundary condition. The Floquet operator $\hat{U}(\lambda)$ becomes diagonal in representation of quasimomentum $k$ of the physical dimension.
We choose the lattice constant to be unity such that
$k\in[-\pi,\pi)$.
Directly solving
\begin{equation}
\hat{U}(k,\lambda) |\psi_j(k,\lambda)\rangle = e^{-iE_j(k,\lambda)} |\psi_j(k,\lambda)\rangle,
\end{equation}
we obtain band eigenstates $|\psi_j(k,\lambda)\rangle$ as well
as the spectrum $E_j(k,\lambda)$ as a function of
$k$, $\lambda$ and the band index $j$.
As a representative example, Fig.~\ref{Fig-Bulk-Spectrum} depicts the obtained three Floquet bands for $J=1$
and $V=-0.5$ (similar behavior can be observed for rather arbitrary values of $J$ and $V$). Two important observations can be made. First, at any individual $\lambda$,
all the three bands are well gapped (hence adiabatic theorem applies).
Second, there is no spectral gap at any given quasienergy $E$.
This is so because at $\lambda=\pm \pi$ the top band smoothly connects the middle band and the middle band smoothly connects the bottom band.  We hence identify
the phase shown in Fig.~\ref{Fig-Bulk-Spectrum} as
a Floquet semimetal with a vanishing indirect gap (viewing $\lambda$ as quasimomentum of a synthetic dimension).
Second, tracking instantaneous $E_j(k,\lambda)$ with $\lambda$ for a complete cycle of $\lambda\rightarrow\lambda+2\pi$, it is seen that
each individual Floquet band as a whole undergoes an exchange with other Floquet bands.
That is, after one adiabatic cycle, though $\hat{U}(k,\lambda)$ has returned to itself,
an eigenstate $|\psi_j(k,\lambda)\rangle$ starting at any $k$ and $\lambda$ will adiabatically evolve to $|\psi_{j'}(k,\lambda)\rangle$, with $j\ne j'$.
In other words, eigenstates of $\hat{U}(k,\lambda)$ from one arbitrary band
adiabatically following $\lambda$
will {\it collectively} evolve to a different band.
This hence generalizes the concept of exotic quantum
holonomy from individual states to Floquet bands as a whole.  Fig.~\ref{Fig-Bulk-Spectrum} also indicates that,  for any band eigenstate to go back to itself (up to a phase), one at least needs to execute three complete cycles in $\lambda$, i.e., $\lambda\rightarrow \lambda+ 6\pi$.

%

\begin{figure} 
\begin{center}
\resizebox{0.4\textwidth}{!}{%
  \includegraphics[trim=0.6cm 1.2cm 0.5cm 10.5cm, clip=true, height=!,width=15cm]{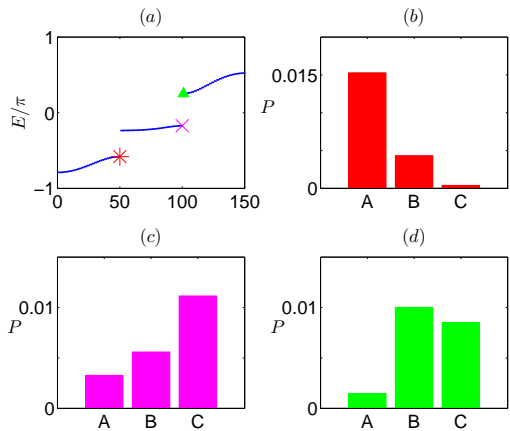}
}
\end{center}
\caption{(color online) Pumping of a bulk eigenstate over $3$ adiabatic
cycles, with $J=1$, $V=-0.5$ and 150 lattice sites.
$(a)$ Quasienergy in ascending order for $\lambda=-\frac{\pi}{2}$ under periodic boundary condition. The $x$-axis represents the index of the states.  The~(red) star denotes the initial state, (magenta) cross or ~(green)
triangle denotes the state
after one or two adiabatic cycles.
Panels $(b)$, $(c)$, and $(d)$ show
sublattice populations for the initial state, state after one
and two adiabatic cycles.}
\label{Fig-Bulk-State-Pump}
\end{figure}

%

\section{Adiabatic dynamics of band eigenstates}
The Floquet-band holonomy may be detected by imaging the exchange of populations within one unit cell.
If the system depicted in Fig.~\ref{Fig-Bulk-Spectrum} is already prepared in a Floquet band eigenstate, then after one adiabatic cycle $\lambda\rightarrow \lambda+2\pi$, the resulting state should be orthogonal to the initial state, leading to detectable population transfer between sublattices.
One simulation result for
$J=1$ and $V=-0.5$ is shown in Fig.~\ref{Fig-Bulk-State-Pump}.
The initial state is located on the bottom Floquet band~(red star) in
Fig.~\ref{Fig-Bulk-State-Pump}$(a)$. Its populations $P_{{\rm A}}$, $P_{{\rm B}}$ and $P_{{\rm C}}$
on sublattice sites A, B, and C are shown in Fig.~\ref{Fig-Bulk-State-Pump}$(b)$, with $P_{{\rm A}}>P_{{\rm B}}>P_{{\rm C}}$. After one adiabatic cycle, the system
evolves to the middle band~[magenta cross in Fig.~\ref{Fig-Bulk-State-Pump}$(a)$],
but now with $P_{{\rm C}}>P_{{\rm B}}>P_{{\rm A}}$ shown in Fig.~\ref{Fig-Bulk-State-Pump}$(c)$.
Consider a second adiabatic cycle, the system evolves to the top
band~[green triangle in Fig.~\ref{Fig-Bulk-State-Pump}$(a)$] and the associated populations
are shown in Fig.~\ref{Fig-Bulk-State-Pump}$(d)$, with $P_{{\rm B}}>P_{{\rm C}}>P_{{\rm A}}$.
Only after a third adiabatic cycle, can all the sublattice populations
return exactly to their initial values.

\begin{figure} 
\begin{center}
\resizebox{0.35\textwidth}{!}{%
  \includegraphics[trim=1.0cm 0.45cm 0.8cm 11.5cm, clip=true, height=!,width=15cm]{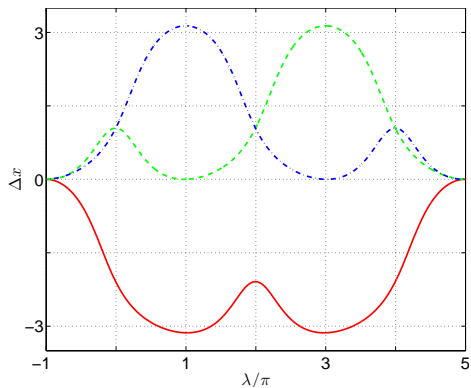}
}
\end{center}
\caption{(color online) Adiabatic pumping of Wannier states over $3$ adiabatic
cycles, with $J=1$ and $V=-0.5$. The wavepacket center
prepared on the bottom, middle and top Floquet
bands in Fig.~\ref{Fig-Bulk-Spectrum} at $\lambda=-\pi$ follows
the red solid, green dashed and blue dashed-dot curves. Each adiabatic cycle
lasts for $4000$ driving periods.}
\label{Fig-Wannier-Pump}
\end{figure}

%

\section{Wannier state pumping}
To further demonstrate Floquet-band holonomy, we consider geometrical adiabatic pumping
starting from band Wannier states~\cite{ThoulessPRB1983,BohmGP2003}.
The induced polarization change along the physical dimension~(called $x$)
can be connected with an integral of the Berry curvature
of a Floquet band~\cite{DerekPRL2012,WangPRB2015,LuPRL2016}. Here, a band
Wannier state at $\lambda=-\pi$ is prepared by uniformly superposing Bloch states with $k=-\pi$
to $k=\pi$ from one chosen band. We then slowly increase $\lambda$.
{Because the system is gapless, the associated Berry curvature integral of any Floquet band is not expected to be an integer.}
Indeed, as seen from the
dynamics of wave packet center in Fig.~\ref{Fig-Wannier-Pump},
the polarization change over one adiabatic cycle
is not quantized, in full agreement with a direct calculation integrating
the Berry curvature. Interestingly, as shown in Fig.~\ref{Fig-Wannier-Pump},
no matter which band the initial Wannier state starts from,
the pumping dynamics of the second or third cycles is different from
that of the first cycle, a clear indicator of
Floquet-band holonomy. Take the bottom band case as an example.
During the first cycle~($\lambda=-\pi\rightarrow\pi$), the wave packet center has always moved towards the $-x$ direction. In the second cycle~($\lambda=\pi\rightarrow3\pi$), there is clearly
an oscillation. The wave packet center returns to its
initial position only after completing the third cycle. Because originally the Wannier
state pumping dynamics is a manifestation of Berry-phase holonomy,
the dynamics we present here reflects an interplay
of Berry-phase holonomy and Floquet-band holonomy. Recent realizations of Thouless's adiabatic
pumping in
optical lattices~\cite{LohseNPS2016,NakajimaNPS2016} suggest that the simulation
results reported here are already within reach of today's experiments.

\begin{figure} 
\begin{center}
\resizebox{0.45\textwidth}{!}{%
  \includegraphics[trim=0.6cm 0.55cm 0.5cm 10.6cm, clip=true, height=!,width=15cm]{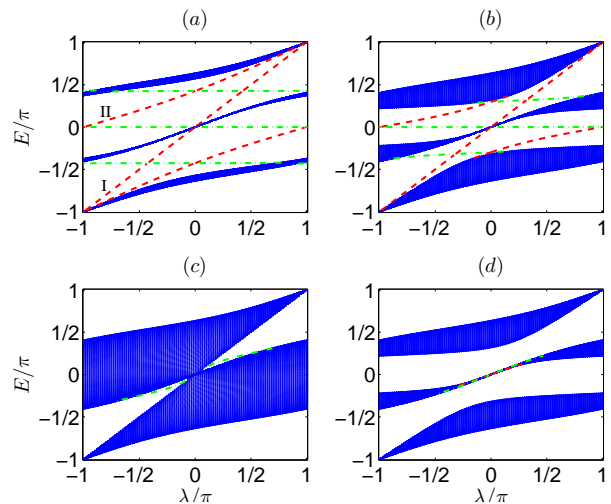}
}
\end{center}
\caption{(color online) Floquet spectrum under open boundary conditions,
The lattice contains $60$ unit cells. Blue dots, red dashed line and green
dashed-dot line denote bulk states, left and right edge modes, with $J=1$.
 $(a)$ $V=-0.9$, with
two types of anomalous chiral edge modes marked by I and II; $(b)$
$V=-0.5$; $(c)$ $V=0$; and $(d)$ $V=0.5$.}
\label{Fig-Spectrum-OBC}
\end{figure}

\section{Anomalous chiral edge modes}
To further investigate
the above-described
Floquet semimetal, we diagonalize
$\hat{U}(\lambda)$ for open lattices.
The associated quasienergy spectrum for different values of $V$ and fixed $J=1$
are shown in Fig.~\ref{Fig-Spectrum-OBC}, with bulk delocalized states represented by shaded areas.
States sufficiently localized at the lattice edges are represented by lines.
We find that anomalous edge modes~(red dashed lines) can wind around the entire quasienergy Brillouin zone $[-\pi, \pi)$ for $V<0$~[see Fig.~\ref{Fig-Spectrum-OBC}$(a)$ and
Fig.~\ref{Fig-Spectrum-OBC}$(b)$], whereas other edge modes~(green dashed-dot lines) do not have this  behavior.  These anomalous edge modes, called
winding edge modes below, are remarkable in at least three aspects. First, they appear only at one end of the open lattice, whereas the other end accommodates non-winding edge modes. Second, Fig.~\ref{Fig-Spectrum-OBC}$(a)$ indicates that there may or may not be Floquet edge mode holonomy.
In particular, type-I edge mode~[marked by ``I" in Fig.~\ref{Fig-Spectrum-OBC}$(a)$] winds around the quasienergy Brillouin zone and returns to itself with $\lambda\rightarrow \lambda+2\pi$; by contrast, type-II edge mode~[marked by ``II" in Fig.~\ref{Fig-Spectrum-OBC}$(a)$] reaches an orthogonal state over one adiabatic cycle and it needs two adiabatic cycles to return to itself. Thus, both type-I and type-II edge modes differ in holonomy behavior
from all bulk states that need three cycles to return to themselves.   Third, it is possible for both type-I and type-II edge modes to be well separated from bulk states for the entire Brillouin zone of $\lambda$. However, type-I edge modes are more robust in this aspect. In particular, as $V$ increases in
Fig.~\ref{Fig-Spectrum-OBC}$(b)$, it is seen that this noteworthy separation from the bulk
persists for type-I edge modes, whereas type-II edge modes start
to merge into the bulk once $V$ is outside the range of $-1.7\lesssim V\lesssim-0.8$.
 A careful inspection accounts for the robustness of type-I edge modes. That is,
type-I edge modes achieve its winding by crossing the middle Floquet band at $\lambda=0$, whose bandwidth turns out to be zero!
In other words, such a crossing with a bulk band is ``safe" because the bulk states tend to be localized and are in general far away from the edges.

We now turn to the case of $V=0$ in Fig.~\ref{Fig-Spectrum-OBC}$(c)$. There the three Floquet bands collapse to two and the above-described anomalous edge modes disappear. This signals a topological phase transition. Indeed, beyond this critical point~[illustrated in Fig.~\ref{Fig-Spectrum-OBC}$(d)$], for arbitrary positive $V$, these anomalous edge modes can no longer be found.
Note in passing that one of the non-winding edge modes seen in Fig.~\ref{Fig-Spectrum-OBC}$(a)$ and Fig.~\ref{Fig-Spectrum-OBC}$(b)$ is a zero-quasienergy mode. Other non-winding edge states localized on the same edge are not zero modes, but also have insignificant dispersion.  The coexistence of anomalous chiral edge modes with the zero modes or with the almost dispersionless edge modes is another interesting feature here.

%

\begin{figure} 
\begin{center}
\resizebox{0.45\textwidth}{!}{%
  \includegraphics[trim=0.6cm 0.5cm 0.5cm 10.6cm, clip=true, height=!,width=15cm]{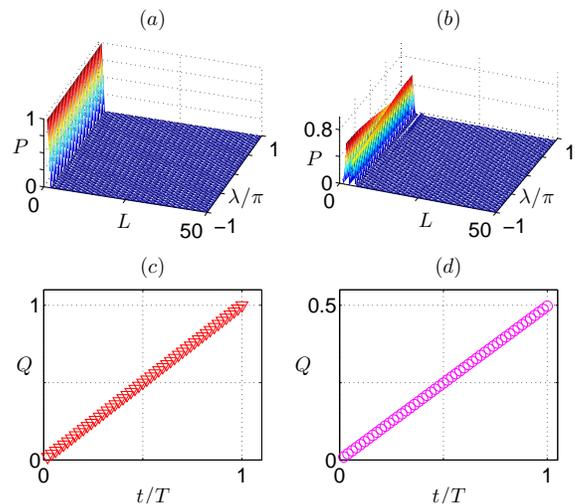}
}
\end{center}
\caption{(color online) Edge state pumping with edge states at different $\lambda$
[shown in Fig.~1$(a)$] uniformly populated.
In simulations the open lattice has $150$ lattice sites.
$(a)$ and $(b)$ show probability distribution $P$ of type-I edge modes
and of the upper branch of type-II edge modes in the first $50$ sites from the left.
$(c)$ and $(d)$ show the time dependence of pumped charge along the synthetic dimension for one driving period for type-I and type-II edge modes.}
\label{Fig-Edge-State-Pump}
\end{figure}

\section{Quantized or half-quantized edge state pumping}
To motivate experimental interest, we now illuminate the implication
of the winding edge modes for charge pumping along the synthetic dimension.
In particular, for either type-I or type-II edge modes, if they are well separated from the bulk for arbitrary $\lambda$, then all such edge states
may be populated without exciting bulk states. Consider then an initial state that uniformly populates all the edge modes for the entire Brillouin zone of $\lambda$. The pumped charge
$Q$ in the synthetic dimension over one driving period is then given by:
\begin{equation}
Q=\sum_{l}n_{l}\int_{0}^{T}dt\langle\psi_{l}(t)|\hat{v}_{\lambda}(t)|\psi_{l}(t)\rangle,
\label{eq:PumpQ}
\end{equation} 
where the group velocity operator $\hat{v}_{\lambda}(t)=\partial{\hat{H}(t)}/\partial{\lambda}$,
$n_{l}$ is the occupation of edge state $|\psi_{l}(0)\rangle$ at $t=0$,
and $|\psi_{l}(t)\rangle$ is the time-evolving state starting from edge mode $|\psi_{l}(0)\rangle$.
Using
\begin{equation}
[\partial_{\lambda}\hat{H}(t)]|\psi_{l}(t)\rangle=
\partial_{\lambda}\left[\hat{H}(t)|\psi_{l}(t)\rangle\right]
-\hat{H}(t)[\partial_{\lambda}|\psi_{l}(t)\rangle]
\end{equation}
 and performing
a time-integration, one has
\begin{equation}
Q=\sum_{l}n_{l} \frac{\partial E^{c}_{l}}{\partial \lambda}.
\end{equation}
where $E^{c}_{l}$ is the quasienergy of edge mode $l$. Now if all type-I edge modes are uniformly populated, then the
summation becomes
\begin{equation}
Q=\frac{1}{2\pi}\int_{-\pi}^{\pi} \frac{\partial E^{c}}{\partial \lambda} d\lambda
=\frac{E^c(\lambda=\pi)-E^c(\lambda=-\pi)}{2\pi},
\end{equation}
which equals precisely $unity$, the winding number of
type-I edge modes. On the other hand, type-II edge modes have Floquet holonomy~(different from the bulk). They need two adiabatic cycles to return to themselves and hence
their effective winding number is $1/2$.  As such,
if only one (e.g. upper) branch of type-II edge modes is uniformly filled,
the pumped charge would be $Q=1/2$.
Our simulation results in Fig.~\ref{Fig-Edge-State-Pump} confirm these insights.

Specifically, Fig.~\ref{Fig-Edge-State-Pump}$(a)$ and \ref{Fig-Edge-State-Pump}$(b)$ present lattice populations of type-I and type-II edge modes
taken from Fig.~\ref{Fig-Spectrum-OBC}$(a)$. It is seen that
these states are always highly localized along the physical dimension.
This feature is true for all values of $\lambda$, even at
$\lambda=0,\pm\pi$ where the edge modes have similar quasienergy values as bulk states.
In Fig.~\ref{Fig-Edge-State-Pump}$(c)$ and \ref{Fig-Edge-State-Pump}$(d)$, we present
the accumulated edge current over a period $T$ by a direct use of Eq.~(\ref{eq:PumpQ}).
It is indeed seen that the pumping
is quantized for type-I edge modes and half quantized for type-II edge modes.
In actual experiments, one may simply start with one single populated edge mode and then tune $\lambda$
``on the fly". The clear separation of the edge modes (type I or II) from  bulk states
guarantees that, one can tune $\lambda\rightarrow \lambda+2\pi$
at a relatively fast rate (e.g., within tens of driving periods) and yet the system
can follow instantaneous edge modes as $\lambda$ varies (hence maintaining the chirality).
As verified by our numerical experiments,
the number of pumped charge per period in this protocol is also quantized or half-quantized.

\section{Conclusions}
We have proposed a Floquet semimetal, with
Floquet-band holonomy and anomalous winding edge modes. Winding edge modes can be always well separated from the bulk, for arbitrary quasimomentum values along a transverse~(synthetic) dimension.
The holonomy behavior of the winding edge states is counter-intuitively different from bulk band states.
All these results are of experimental interest.
We finally make a connection
between this work with the discovery of AFAI~\cite{TitumPRX2016} where quantized edge state pumping was first investigated. The Floquet semimetal studied here clearly shows that,
even in the absence of any disorder, it is possible to have quantized or even half-quantized edge state pumping.

The intriguing interplay between Floquet topological phases and Floquet
holonomy deserves further investigations. Theoretically, whether
the chiral edge modes in the semimetal phase could be fully characterized
by a winding number should be further studied. A systematic topological classification
of our system is also an interesting topic. Experimentally, due to the successful
engineering of several Floquet topological phases in cold-atom and
photonic systems~\cite{RechtsmanNat2013,JotzuNat2014}, we expect
the proposed model here to be within reach of current quantum simulation
technologies. Moreover, thanks to the remarkable separation between edge
modes and bulk states, together with the already existing experimental setups for
single bulk-state pump~\cite{LuPRL2016}, Thouless pump~\cite{LohseNPS2016,NakajimaNPS2016} and
edge state visualization~\cite{HafeziNatPhots2013,AtalaNPs2014,ManciniSci2015,StuhlSci2015},
it should be possible to detect the exotic quantum holonomy
of bulk quasimomentum states, Wannier states and chiral edge modes
in optical or photonic lattices.

\begin{acknowledgments} {\bf Acknowledgments:} We thank Hailong Wang and Derek Y.H. Ho for helpful discussions. This work
is supported by Singapore Ministry of Education Academic Research
Fund Tier I~(WBS No. R-144-000-353-112).
\end{acknowledgments}


\end{document}